# Massless and massive particle-in-a-box states in single-and bi-layer graphene

Sungjae Cho and Michael S. Fuhrer

*Department of Physics and Center for Nanophysics and Advanced Materials, University of Maryland, College Park, Maryland 20742, USA*

Electron transport through short, phase-coherent metal-graphene-metal devices occurs via resonant transmission through particle-in-a-box-like states defined by the atomically-sharp metal leads. we study the spectrum of particle-in-a-box states for single- and bi-layer graphene, corresponding to massless and massive two-dimensional (2d) Fermions.The density of states $D$ as a function of particle number $n$ shows the expected relations $D(n) \sim n^{1/2}$ for massless 2d Fermions (electrons in single-layer graphene) and $D(n) \sim$ constant for massive 2d Fermions (electrons in bi-layer graphene). The single parameters of the massless and massive dispersion relations are found, Fermi velocity $v_F = 1.1 \times 10^6$m/s and effective mass $m^* = 0.032\ m_e$, where $m_e$ is the electron mass, in excellent agreement with theoretical expectations.

The problem of a wavelike particle confined to a hard-walled box is one of the most basic problems in quantum mechanics. The spectra of the particle-in-a-box are strikingly different for massive and massless particles: massless particles (e.g. photons) have energies which depend linearly on quantum number, while the energies of massive particles (e.g. free electrons) depend quadratically on quantum number. For Fermions in two dimensions (2d) this leads to a density of single-particle states $D \equiv dn/dE$, where $E$ is the particle energy and $n$ the particle density, which varies as the square root of particle density, i.e. $D(n) \sim n^{1/2}$, for massless particles, and is independent of particle density for massive particles, i.e. $D(n) \sim$ constant. Here we show that mesoscopic, ballistic[1] single-layer[2,3] and bi-layer[4,5] metal-graphene-metal devices act as Fabry-Perot cavities for electrons confined between the atomically-sharp partially-reflective metal leads. Electronic conduction occurs through resonant states of the Fabry-Pérot cavity, which are exactly analogous to the particle-in-a-box states of an electron confined by perfectly reflective walls. $D(n)$ is measured, and the expected dependences on particle number are verified: $D \sim n^{1/2}$ for massless particles in single-layer graphene, and $D \sim$ constant for massive particles in bi-layer graphene. $D(n)$ is used to extract the single constants in the dispersion relations: the Fermi velocity $v_F = 1.09 \times 10^6$ m/s for massless particles in single-layer graphene[2,3], and the effective mass $m^* = 0.032\ m_e$, where $m_e$ is the electron mass, for massive particles in bilayer graphene[4,10] in excellent agreement with theoretical expectations[5,10,13,21,22] and other experimental results[2,3,19,20,23].

We first review the results of the two-dimension particle-in-a-box problem. Figs. 1A and 1B, illustrate the massless and massive dispersion relations respectively:

$E = \hbar v_F |\boldsymbol{k}|$ (massless) (1a) $E = \frac{\hbar^2 k^2}{2m^*}$ (massive)

(1b)

where $\hbar$ = $h/2\pi$, and $h$ is Planck's constant; each dispersion relation is characterized by a single parameter, $m^*$ for the massive dispersion, and $v_F$ for the massless dispersion. For particles confined to a 2-dimensional box of width $W$ and length $L$ the hard-wall boundary condition quantizes the wavevector $\mathbf{k} = (k_x, k_y) = \left( p\frac{\pi}{L}, q\frac{\pi}{W} \right)$ resulting in two positive quantum numbers $p$, $q$. Figs. 1C and 1D illustrate this quantization, where each point represents an allowed wavevector. Then the energies in terms of quantum numbers are given by the familiar relations $E = \hbar v_F \pi \sqrt{\left(\frac{p}{W}\right)^2 + \left(\frac{q}{L}\right)^2}$ (massless) and $E = \frac{h^2}{8m}\left\{\left(\frac{p}{W}\right)^2 + \left(\frac{q}{L}\right)^2\right\}$ (massive). For Fermions at zero temperature, the occupancy of particle-in-a-box state will be the degeneracy of individual states $g$ for state of energy $E < E_F$ (the Fermi energy) and zero for states $E > E_F$. The number of states with $E < E_F$ is given by $N = g k_F^2 WL / 4\pi$, where $k_F \equiv k(E_F)$ is the Fermi wavevector. Figs. 1C and 1D illustrate the occupied states included for equally-spaced values of $E_F$, and Figs. 1E and 1F show the energies of the particle-in-a-box states as a function of particle number $N$ for massless and massive 2d Fermions respectively. The linear and square-root dependences of $E(N)$ for massless and massive 2d

Fermions respectively are evident in Figs. 1E and 1F. Using the areal density of particles $n = \frac{N}{WL} = \frac{g k_F^2}{4\pi}$ we then have the following relations for the dependences of the Fermi energy $E_F$ and density of states $D$ on density:

$$E_F = \hbar v_F \sqrt{\pi n} \text{ (massless)} \quad \text{(2a)} \qquad E_F = \frac{\pi \hbar^2 n}{2m^*} \text{ (massive)} \quad , \qquad \text{(2b)}$$

$$D = \frac{1}{\hbar v_F}\sqrt{\frac{gn}{\pi}} \text{ (massless)} \quad \text{(3a)} \qquad D = \frac{g m^*}{2\pi\hbar^2} \text{ (massive).} \qquad \text{(3b)}$$

Thus the measurement of $D$ as a function of $n$ distinguishes massive and massless particles, and (given knowledge of the degeneracy $g$) also determines the constants of the dispersion relations $v_F$ and $m^*$.

Single- and bi-layer graphene may be used to realize the dispersion relations in Equations 3a and 3b as follows. Single-layer graphene is well-described by a tight-binding model considering only the π-orbitals at each atomic site. At zero doping, the π and π* bands meet at two points in the Brillouin zone with wavevector **K**. This crossing is preserved as long as the two atoms A and B in the unit cell are equivalent. Taking $E(\mathbf{K}) = 0$, and measuring $k$ away from the **K** point, the band structure is well-approximated by Eqn. 1a, with $v_F = \left(\sqrt{3}/2\right) a \gamma_0 / \hbar \approx 1.0 \times 10^6$ m/s where $a$ = 2.46 Å is the graphene lattice constant and $\gamma_0 \approx 3.16$ eV [9] is the nearest-neighbor hopping parameter. In Bernal-stacked bi-layer graphene [4,5,10,21,22] atom A in one layer is stacked above atom B' in the 2nd layer, and this A-B' coupling breaks the AB equivalency of the graphene unit cell and results in two bands which may be approximated as hyperbolic: $E_\pm(k) = \pm\left[\sqrt{(\eta v_F k)^2 + \frac{\gamma_1^2}{4}} - \frac{\gamma_1}{2}\right]$ [10], where $\gamma_1 \approx 0.4$ eV[24] is the inter-layer (A-B') hopping parameter. At $k$ = 0 the effective mass is given by $m^* = \gamma_1 / 2v_F^2 \approx 0.03\, m_e$. In both single- and bilayer graphene the degeneracy $g$ = 4, due to the two-fold spin degeneracy and the two-fold valley degeneracy (presence of two **K** points).

We now discuss the graphene samples used in this study. We mechanically exfoliate Kish graphite on 300nm $SiO_2$/Si substrates to obtain single and bilayer graphene[2,3,4]. Single layer graphene is more transparent than two or more layer graphene under optical microscope as seen in Fig. 2. After locating graphene flakes, Cr/Au(5nm/50nm) were thermally deposited for electrical contacts. The channel lengths $L$ for Fabry-Perot interference measurement are 200nm - 300nm and measured by scanning electron microscope. The maximum field-effect mobilities at low temperature estimated from the four -probe resistivity of the adjacent graphene sections are 15,000 $cm^2$/Vs and 4,000 $cm^2$/Vs for single- and bilayer graphene respectively.Figs. 2A and 2B show completed single- and bilayer graphene devices respectively. Electrodes are patterned on each graphene flake to form a large-area Hall-bar arrangement for characterizing the longitudinal and Hall conductivities ($\sigma_{xx}$ and $\sigma_{xy}$) of the sample. In addition, pairs of closely-spaced (150-300 nm) electrodes act as Fabry-Pérot cavities on the same sample.

Figs. 2C and 2D show $\sigma_{xx}$ and $\sigma_{xy}$ for the single- and bi-layer graphene devices shown in Figures 1a and 1b, respectively, measured in high magnetic field (9 T) and as a function of back-gate voltage $V_g$, which controls the carrier density $n = c_g V_g / e$, where $c_g = 1.1 \times 10^{-8}$ F/m, and $e$ is the electronic charge. The quantized Hall effect (QHE) is evident as plateaux with $\sigma_{xy} = \nu e^2/h$, and corresponding minima in $\sigma_{xx}$. Berry's phases of π and 2π lead to QHE in single- and bi-layer graphene at filling factors $\nu = 4(i+1/2)$ and $4(i+1)$, where $i$ is an integer[refs], thus our observation of the half-integer and full-integer QHE confirms the identification of these samples as single- and bi-layer graphene respectively.

Figs. 2E and 2F show the two-probe conductances as a function of gate voltage $G(V_g)$ for Fabry-Pérot cavities on the single- and bi-layer devices, respectively, at zero magnetic field. We shift the curves horizontally by an amount $V_D$ which we identify as the gate voltage at which the Fermi level lies closest to the Dirac point. The conductance rises away from $V_g - V_D = 0$ as observed by previously[2,3,4]. Small reproducible fluctuations of the conductance with magnitude of order $e^2/h$ can be seen (see insets to Figs. 2E and 2F); these fluctuations are not universal conductance fluctuations (UCF; see Supplemental Information) but, as argued below, result from the interference of ballistic electron waves in the Fabry-Pérot cavity [1].

Figure 3 shows color-scale maps of the differential conductance d$I$/d$V$ as a function of bias voltage $V$ applied between the two electrical contacts and gate voltage $V_g$ for the single- and bi-layer devices shown in Figure 2. A pattern of diagonal lines of increased conductance is evident; this pattern is the signature of Fabry-Pérot interference in a mesoscopic device[1,6]. Neighboring diagonal lines have similar slopes, and diagonal line of similar positive and negative slope are found in each $V_g$ region. Each individual diagonal line results from the enhancement in conductance when a particle-in-a-box resonance, or a group of constructively-interfering resonances, is aligned with the source electrode (+$V$) or drain electrode (-$V$); the symmetry about $V = 0$ reflects the source-drain symmetry of the device. Note that the pattern is inconsistent with Coulomb blockade; there are no diamond-shaped low conductance regions around $V = 0$, and the overall conductance >> $e^2/h$ excludes Coulomb blockade.

Resonant transmission through a Fabry-Pérot cavity has been reported previously for carbon nanotubes (CNTs) [6,11] and graphene [1]. In the case of CNTs, there is a single path length $L$ connecting the electrodes, and the resonances are evenly spaced in $V$ and $V_g$ (see Methods). In graphene [1], the resonances are randomly spaced, which may result from a spread of path lengths due to non-parallel electrodes or electron paths which are not perpendicular to the electrode-graphene interfaces. However, important information can be gained by analyzing the *slope* $\Delta V/\Delta V_g$ of the resonant lines in Fig. 3. Briefly, the slope measures the change in energy $\Delta E = e\Delta V/2$ (the factor of two results from the potentials $+\Delta V/2$ and $-\Delta V/2$ applied to the two electrodes relative to the graphene in a ballistic device) of the resonance as the particle number $dn = c_g\Delta V_g/e$ is changed. The slope is then equal to $\Delta V/\Delta V_g = (2c_g/e^2)\Delta E/\Delta n = (2c_g/e^2)D^{-1}$; i.e. the slope is inversely proportional to the density of states (see Methods for a more rigorous derivation of the same result). From Eqns. 3a and 3b above, we expect that $\Delta V/\Delta V_g \sim n^{-1/2} \sim |V_g - V_D|^{-1/2}$ for the single-layer (massless dispersion) sample, and $\Delta V/\Delta V_g$ ~ *constant* for the bi-layer (massive dispersion) sample. Figs. 3A-C show that the slope indeed varies significantly with gate voltage (electron density) for the single-layer graphene sample, with the highest slope occurring near $V_g - V_D = 0$. The slope is nearly constant in the bi-layer graphene sample (Figs. 3D-F).

Fig. 4 plots the density of states $D = (e^2/2c_g)(\Delta V/\Delta V_g)^{-1}$ for the single- and bi-layer graphene samples extracted from Figs. 3A-F and additional data (not shown) as a function of electron density $n = c_g(V_g - V_D)/e$. Data from an additional single-layer sample is also shown (see Supplementary Information). Solid lines are fits to Eqns. 3a and 3b for the single-layer and bi-layer data respectively. The expected dependences on particle number are verified: $D \sim n^{1/2}$ for massless particles in single-layer graphene (Eqn. 3a), and $D$ ~ *constant* for massive particles in bi-layer graphene (Eqn. 3b). Only a single fitting parameter is used in each fit, $v_F = (1.09 \pm 0.01) \times 10^6 m/s$ for massless particles in single-layer graphene and $m^* = (0.0315 \pm 0.0001) m_e$ for massive particles in bi-layer graphene. As discussed in detail below, the parameters are in excellent agreement with theoretical and other experimental results.

We now discuss the detailed dependence of the density of states on particle number in single- and bi-layer graphene, and the implications of the results for understanding the electronic structure of these materials. From the fit to Eqn. 3a in Figure 4, we determine a Fermi velocity for single-layer graphene of $v_F = (1.09 \pm 0.01) \times 10^6 m/s$. A tight-binding model of graphene [12] gives $v_F = (\sqrt{3}/2) a\gamma_0 / \hbar \approx 1.0 \times 10^6$ m/s where $a$ = 2.46 Å is the graphene lattice constant and $\gamma_0 \approx 3.16$ eV [9] is the nearest-neighbor hopping parameter. The inclusion of electron-electron interactions will renormalize the Fermi velocity slightly [13], and the slightly higher $v_F$ observed here is consistent with other experiments on graphene [2,3,14,15].

The density of states in single-layer graphene remains finite as $n \rightarrow 0$ due to charge inhomogeneity caused by charged impurities near the graphene, as has been observed previously. The minimum density of states $D$ on order 2 x $10^{12}$ $eV^{-1}cm^{-2}$ corresponds to a charge density $n$ on order $10^{11}$ $cm^{-2}$, in agreement with theoretical [16] and experimental expectations [8,17] for the minimum charge density at the Dirac point in the presence of charged impurity disorder due to the $SiO_2$ substrate.

We now discuss bilayer graphene. From the fit to Eqn. 3b in Figure 4, we determine $m^* = 0.032m_e$. Assuming $v_F = 1.09 \times 10^6$ m/s we have $\gamma_1 = 0.40$ eV, in excellent agreement with the experimental values for graphite of $0.39 \pm 0.01$ eV [18] and with other experiments on bilayer graphene [19, 20]. Because the bands are not strictly parabolic, the density of states should depend on particle density, increasing with increasing particle density. The hyperbolic nature of the bands becomes important for particle densities roughly greater than $\gamma_1^2/(4\pi\hbar^2 v_F^2) \approx 3 \times 10^{12}$ $cm^{-2}$. Experimentally, we see little variation in the density of states for particle densities up to 6 x $10^{12}$ $cm^{-2}$, indicating a wider range of validity of the parabolic spectrum than expected. We do not currently understand this discrepancy, but we note that electron-electron interactions should again be important, as was pointed out previously in the failure of the single-particle picture to quantitatively explain the cyclotron resonance spectrum in bilayer graphene [20].

In conclusion, we have probed the density of particle-in-a-box states as a function of particle number for massless 2d Fermions (single-layer graphene) and massive 2d Fermions (bi-layer graphene) in phase-coherent measurement. Understanding of coherent transport is an essential step to realize other interesting experiments in graphene such as focusing lens, klein tunneling and graphene superlattice[25,26,27,28]. The density of states varies as the square-root of particle number for massless 2d Fermions, and is constant for massive 2d Fermions. The single parameters in the dispersion relations are extracted; the Fermi velocity $v_F = (1.09 \pm 0.01) \times 10^6 m/s$ for massless particles in single-layer graphene and $m^* = (0.0315 \pm 0.0001) m_e$ for massive particles in bi-layer graphene, in excellent agreement with theoretical expectations and other experimental observations.

**Acknowledgements**

This work has been supported by the U.S. ONR grant N000140610882, NSF grant CCF-06-34321, and the UMD-NSF-MRSEC grant DMR-05-20471. We acknowledge useful discussions with S. Adam and S. Das Sarma.

**Figure Captions**

Fig. 1. (A) Massless dispersion and (B) massive dispersion relations in two dimensions. (C-D) Allowed wavenumbers for particle in a box of aspect ratio *W*/*L* = 1.6. Solid lines are contours of equal energy for massless dispersion relation (C) and massive dispersion relation (D). (E-F) Particle energy as a function of particle number in a box with *W*/*L* = 1.6 for massless dispersion relation (E) and for massive dispersion relation (F).

Fig. 2. (A-B) Optical micrographs of single-layer graphene device (A) and bilayer graphene device (B). (C-D) Longitudinal and Hall conductivity as a function of gate voltage at magnetic field of 9 T and temperature of 1.3 K for single-layer graphene device (C) and bilayer graphene device (D). (E-F) Two-probe conductance as a function of gate voltage at zero magnetic field and temperature of 1.3 K for single-layer graphene device (E) and bilayer graphene device (E)

Fig. 3. Color-scale two-dimensional plots of differential conductance $G = dI/dV$ as a function of bias voltage $V$ and gate voltage $V_g$ measured in single (A-C) and bilayer (D-F) graphene at temperature $T$ = 1.3K. A smooth background conductance was subtracted to enhance the patterns. The sample dimensions are 1.5μm(*W*) x 0.3μm(*L*) for single layer graphene and 4.3μm(*W*) x 0.2μm(*L*) for bilayer graphene. Yellow lines illustrate the slope of the Fabry-Pérot resonances.

Fig. 4. Density of states of single-layer graphene (blue symbols; data from two devices shown) and bilayer graphene (red symbols, data from one device shown) as a function of particle density. Solid lines are fits to Eqn. 3a (blue) with $v_F$ = $1.09 \times 10^6$ m/s and Eqn. 3b (red) with $m^* = 0.032\, m_e$.

Figure 1

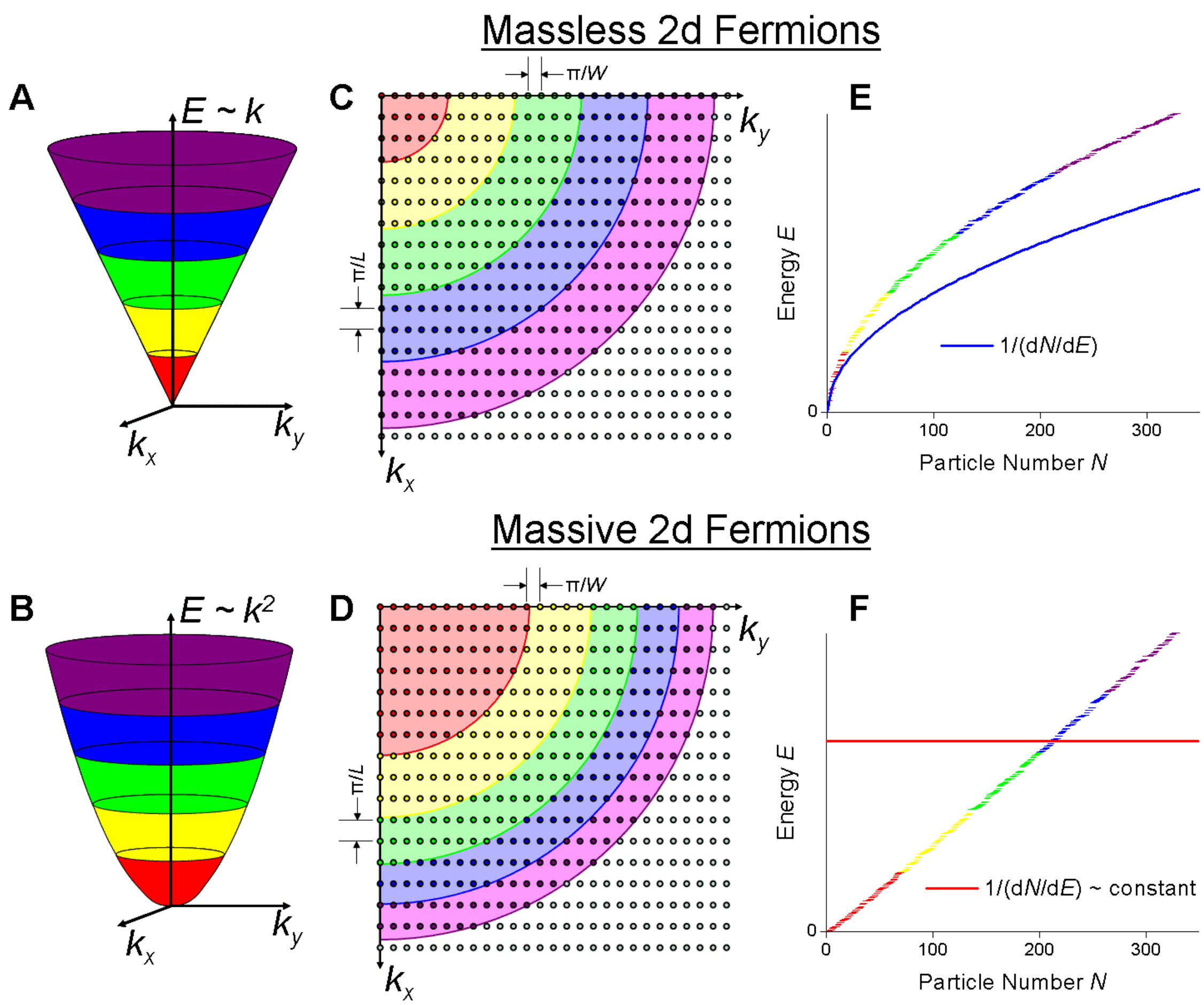

Massless 2d Fermions
A
E ~ k
k_x
k_y
C
π/W
π/L
k_y
k_x
E
Energy E
0
0
100
200
300
Particle Number N
1/(dN/dE)
Massive 2d Fermions
B
E ~ k^2
k_x
k_y
D
π/W
π/L
k_y
k_x
F
Energy E
0
0
100
200
300
Particle Number N
1/(dN/dE) ~ constant

Figure 2

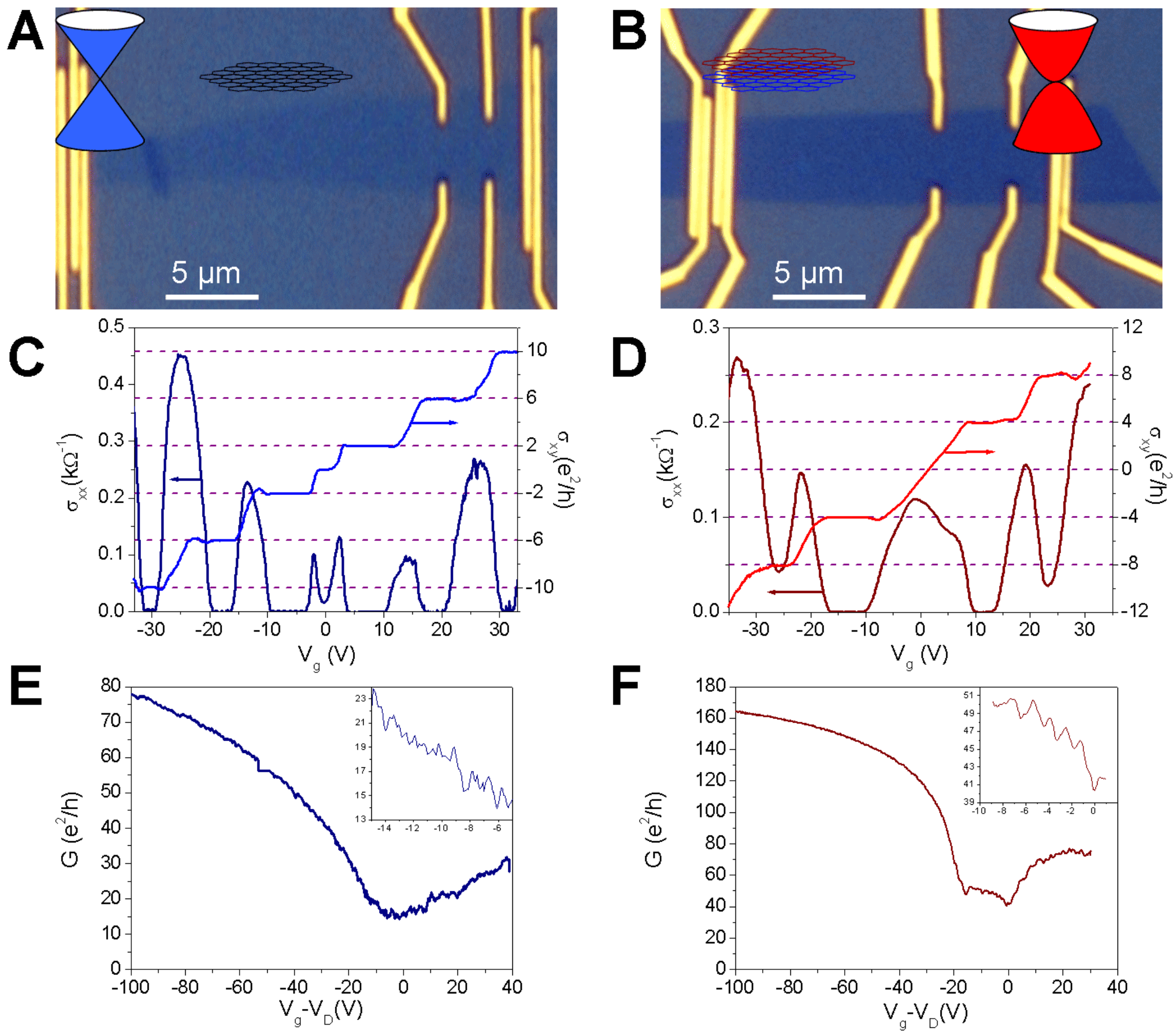

Figure 3

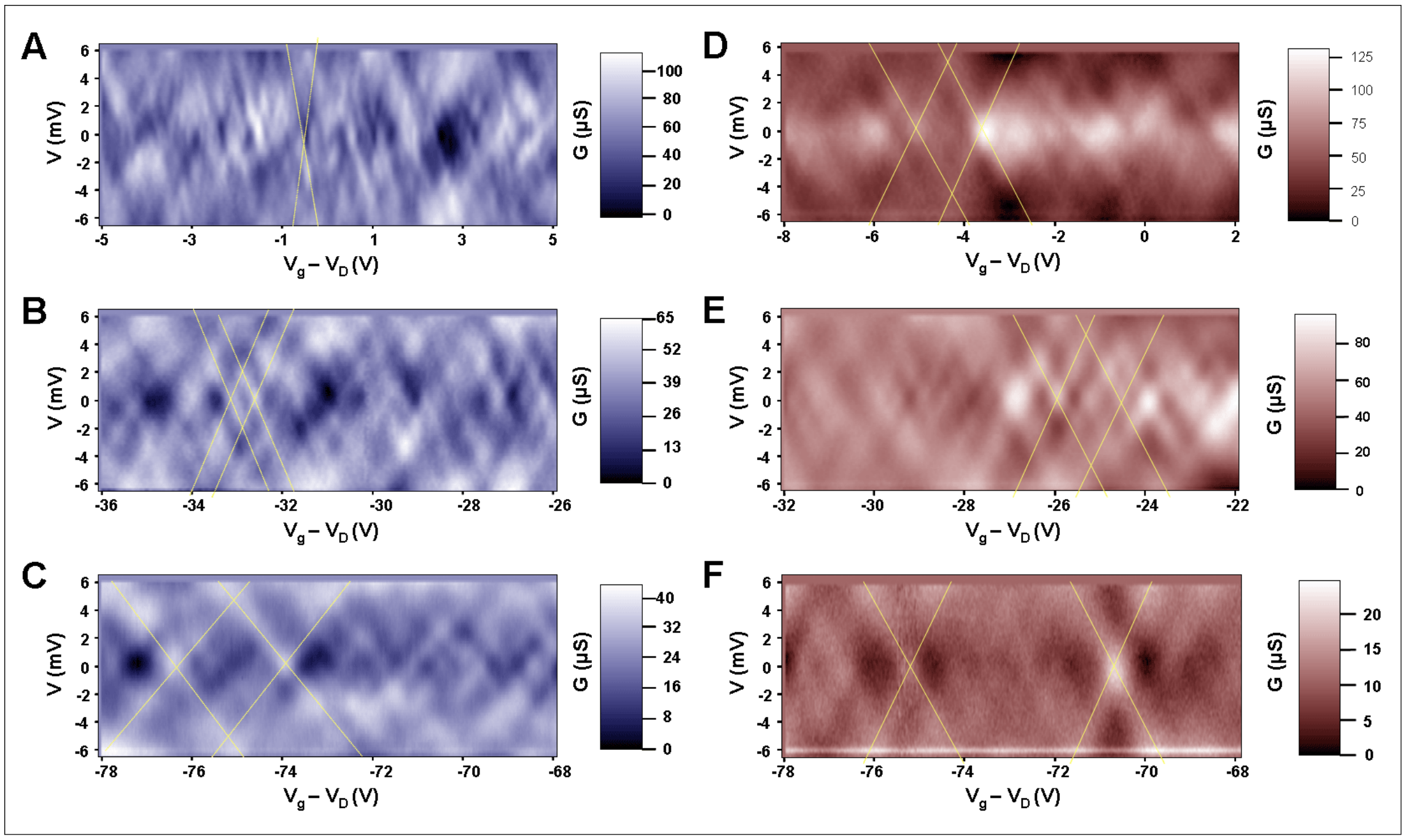

Figure 4

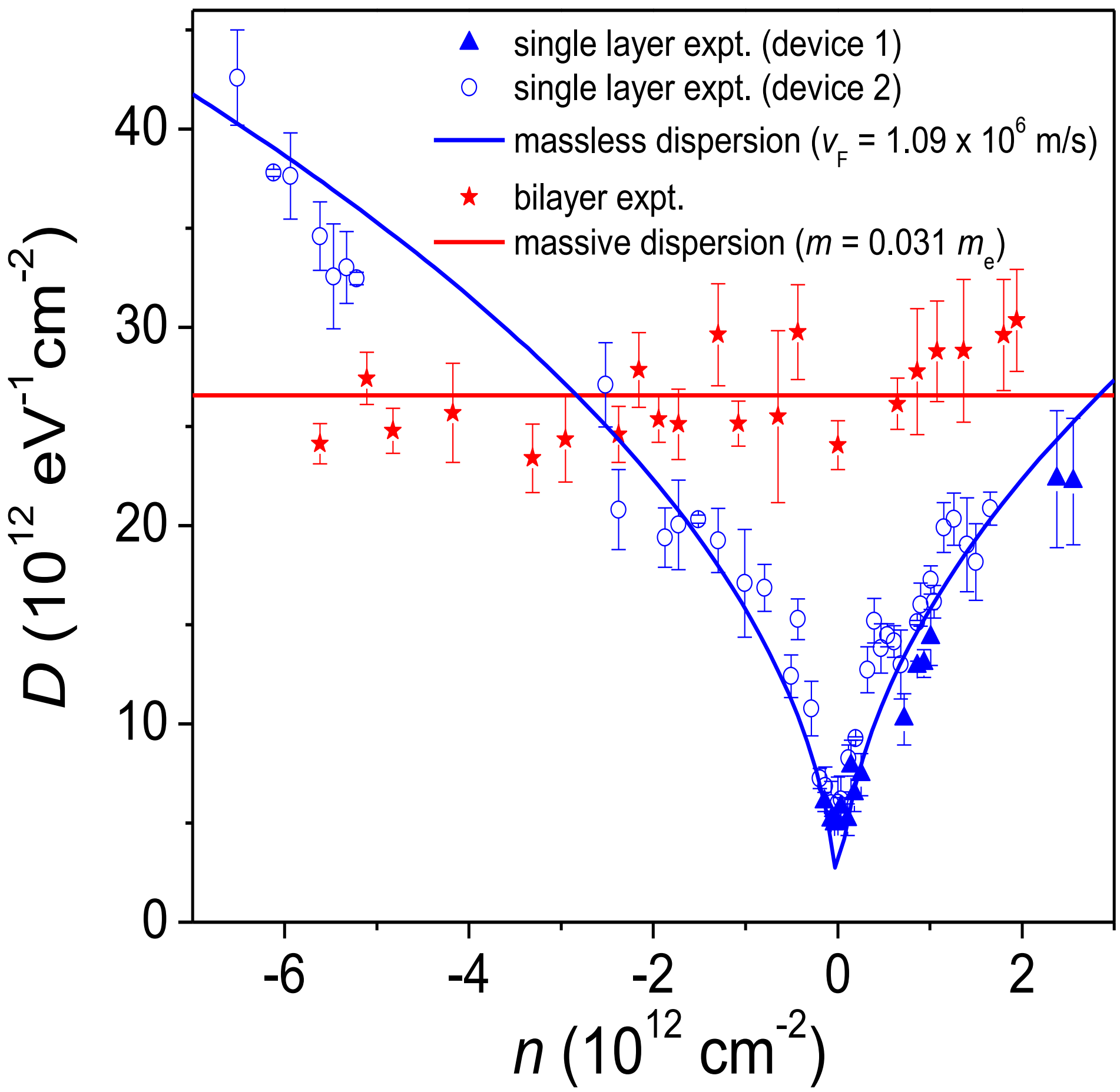
single layer expt. (device 1)
single layer expt. (device 2)
massless dispersion ($v_F$ = 1.09 x $10^6$ m/s)
bilayer expt.
massive dispersion ($m$ = 0.031 $m_e$)
$D$ ($10^{12}$ eV$^{-1}$cm$^{-2}$)
$n$ ($10^{12}$ cm$^{-2}$)
0
10
20
30
40
-6
-4
-2
0
2

**References**

1. Miao, F. *et al*, Phase-coherent transport in graphene quantum billiards. *Science* **317**, 1530(2007)
2. Novoselov, K. S. *et al*, Two-dimensional gas of massless Dirac fermions in graphene. *Nature* **438**, 197(2005)
3. Zhang, Y. *et al*, Experimental observation of the quantum Hall effect and Berry's phase in graphene, *Nature* **438**, 201(2005)
4. Novoselov, K. S. *et al*, Unconventional quantum hall effect and Berry's phase of $2\pi$ in bilayer graphene. *Nature Physics* **2**, 177(2006)
5. Koshino, M. and Ando, T. Transport in bilayer graphene: Calculations within a self-consistent Born approximation. *Phys. Rev. B* **73**, 245403 (2006)
6. Liang, W. *et al*, Fabry-perot interference in a nanotube electron waveguide. *Nature* **411**, 665(2001)
7. Hwang, E.H. et al, Density Dependent Exchange Contribution to $\partial\mu/\partial n$ and Compressibility in Graphene. *Phys. Rev. Letter* **99**, 226801(2007)
8. Martin, J. *et al*, Observation of electron–hole puddles in graphene using a scanning single-electron transistor. *Nature physics* **4**, 144 - 148 (2008)
9. Toy, W. W. et al, Minority carriers in graphite and the *H*-point magnetoreflection spectra, *Phys. Rev. B* **15,** 4077 - 4090 (1977)

10. McCann, E. *et al*, Landau-Level Degeneracy and Quantum Hall Effect in a Graphite Bilayer, *Phys. Rev. Lett.* **96**, 086805 (2006)

11. Cao, Electron transport in very clean, as-grown suspended carbon nanotubes, *Nature Material*s **4**, 745 (2005)

12. Wallace, P. R., The Band Theory of Graphite, *Phys. Rev.* **71**, 622 - 634 (1947)

13. Das Sarma, S. *et al*, Many-body interaction effects in doped and undoped graphene: Fermi liquid versus non-Fermi liquid, *Phys. Rev. B* **75**, 121406(R) (2007)

14. Jiang, Z. *et al*, Infrared Spectroscopy of Landau Levels of Graphene, *Phys. Rev. Lett.* **98**, 197403 (2007)

15. Sadowski, M. L. *et al*, Landau Level Spectroscopy of Ultrathin Graphite Layers, *Phys. Rev. Lett.* **97**, 266405 (2006)

16. Adam, S. *et al*, A self-consistent theory for graphene transport, *PNAS* **104**, 18392 (2007)

17. Chen, J. *et al*, Charged Impurity Scattering in Graphene, *Nature Physics* (2008)

18. Misu, A. *et al*, Near Infrared Reflectivity of Graphite under Hydrostatic Pressure. I. Experiment, *J. Phys. Soc. Jap*. **47**, 199 (1979)

19. Ohta, T. *et al*, Controlling the Electronic Structure of Bilayer Graphene, *Science* **313**, 951 (2006)_

20. Henriksen, E. A. *et al*, Cyclotron Resonance in Bilayer Graphene, *Phys. Rev. Lett*. **100**, 087403 (2008)

21. Nilson, J. *et al*, Electronic Properties of Graphene Multilayers, *Phys. Rev. Lett.* **97**, 266801 (2006)

22. Partoens, B. *et al*, From graphene to graphite: Electronic structure around the *K* point, *Phys. Rev. B* **74**, 075404 (2006)

23. Bostwick, A. *et al*, Quasiparticle dynamics in graphene, *Nature Physics* **3**, 36 - 40 (2007)

24. Dresselhaus, M. S. *et al*, Intercalation compound of graphite, *Advances in Physics* **30**, 139 (1981)

25. Cheianov, V.  *et al*, Selective transmission of Dirac electrons and ballistic magnetoresistance of *n*-*p* junctions in graphene , *Phys. Rev. B* **74**, 041403(R) (2006)

26. Cheianov, V.  *et al*, The focusing of Electron Flow and a Veselago Lens in Graphene *p-n* Junctions, Science **315**, 1252 (2007)

27. Katsnelson, M. *et al*, Chiral tunnelling and the Klein paradox in graphene, *Nature Physics* **2**, 620 - 625 (2006)

28. Park, C. , *et al*, Anisotropic behaviours of massless Dirac fermions in graphene under periodic potentials, *Nature Physics* 4, 213(2008)